\documentclass[envcountsame,runningheads,orivec]{llncs}
\pdfoutput=1
\usepackage{a4}
\usepackage{amsmath}
\usepackage{amssymb}
\usepackage{xspace}
\usepackage{graphicx}
\DeclareFontFamily{U}{bbold}{}
\DeclareFontShape{U}{bbold}{m}{n}
   {  <5> <6> <7> <8> <9> gen * bbold
      <10> <10.95> bbold10
      <12> <14.4> bbold12
      <17.28> <20.74> <24.88> bbold17
   }{}
\DeclareSymbolFont{bbold}{U}{bbold}{m}{n}
\DeclareMathSymbol{\BssD}{3}{bbold}{"01}
\DeclareMathSymbol{\BssS}{3}{bbold}{"06}
\DeclareMathSymbol{\BssP}{3}{bbold}{"05}
\DeclareMathSymbol{\BssReduceq}{3}{bbold}{"3C}

\newcommand{\IN}{\mathbb{N}}
\newcommand{\IR}{\mathbb{R}}
\newcommand{\IH}{\mathbb{H}}
\newcommand{\IL}{\mathbb{L}}
\newcommand{\IM}{\mathbb{M}}

\newcommand{\IZ}{\mathbb{Z}}
\newcommand{\IC}{\mathbb{C}}
\newcommand{\IK}{\mathbb{K}}
\newcommand{\IQ}{\mathbb{Q}}
\newcommand{\IO}{\mathbb{O}}

\newcommand{\IT}{\mathbb{T}}
\newcommand{\IS}{\mathbb{S}}
\newcommand{\ISD}{\IS^{\text{d}}}
\newcommand{\ISO}{\IS^{\text{o}}}
\newcommand{\II}{\mathbb{I}}
\newcommand{\IU}{\mathbb{U}}
\newcommand{\Kol}{K}
\newcommand{\KolO}{\Kol^{\text{o}}}
\newcommand{\KolS}{\Kol^{\text{s}}}
\newcommand{\KolD}{\Kol^{\text{d}}}
\newcommand{\KOL}{\IK}
\newcommand{\KOLO}{\KOL^{\text{o}}}
\newcommand{\KOLS}{\KOL^{\text{s}}}
\newcommand{\KOLD}{\KOL^{\text{d}}}
\newcommand{\dom}{\operatorname{dom}}
\newcommand{\length}{\operatorname{length}}
\newcommand{\size}{\operatorname{size}}
\newcommand{\sign}{\operatorname{sign}}
\newcommand{\trdeg}{\operatorname{trdeg}}
\newcommand{\Card}{\operatorname{Card}}
\newcommand{\calO}{\mathcal{O}}
\newcommand{\calS}{\mathcal{S}}
\newcommand{\calX}{\mathcal{X}}
\newcommand{\calY}{\mathcal{Y}}
\newcommand{\person}[1]{\textsc{#1}}
\newcommand{\COMMENTED}[1]{}

\newcommand{\mycite}[2]{\cite[\textsc{#1}]{#2}}
\newcommand{\BCSS}{BSS\xspace}
\newcommand{\BSS}{BSS\xspace}

\newtheorem{observation}[theorem]{Observation}
\spnewtheorem{fact}[theorem]{Fact}{\bfseries}{\itshape}
\newcommand{\range}{\operatorname{range}}
\newcommand{\lspan}{\operatorname{lspan}}
\def\vecbf#1{\mathchoice{\mbox{\boldmath$\displaystyle#1$}}
{\mbox{\boldmath$\textstyle#1$}}
{\mbox{\boldmath$\scriptstyle#1$}}
{\mbox{\boldmath$\scriptscriptstyle#1$}}}
\begin{document}
\title{Kolmogorov Complexity Theory over the Reals}
\titlerunning{Kolmogorov Complexity Theory over the Reals}
\authorrunning{M.~Ziegler and W.M.~Koolen}
\tocauthor{M.~Ziegler and W.M.~Koolen}
\author{Martin Ziegler\inst{1}\thanks{supported by the German Research
Foundation (DFG) with project \texttt{Zi\,1009/1-1}}
\and Wouter M. Koolen\inst{2}}
\institute{
University of Paderborn, Germany; \email{ziegler@upb.de}
\and
CWI, Amsterdam, The Netherlands; \email{wmkoolen@cwi.nl}}

\date{}
\thispagestyle{empty}
\maketitle
\def\thefootnote{\fnsymbol{footnote}}\addtocounter{footnote}{1}
\begin{abstract}
Kolmogorov Complexity constitutes an integral part of 
computability theory, information theory, and computational 
complexity theory---in the discrete setting of bits and Turing machines.
Over real numbers, on the other hand,
the \BSS-machine (aka real-RAM) has been established as a
major model of computation.
This real realm has turned out to exhibit natural counterparts 
to many notions and results in classical complexity and recursion 
theory; although usually with considerably different proofs.
The present work investigates similarities and differences 
between discrete and real Kolmogorov Complexity as introduced
by Monta\~{n}a and Pardo (1998).
\end{abstract}
\section{Introduction}
It is fair to call Andrey Kolmogorov 
one of the founders of Algorithmic Information Theory.
Central to this field is a formal notion of information
content of a fixed finite binary string $\bar x\in\{0,1\}^*$:
For a (not necessarily prefix)
universal machine $U$ 
let $\Kol_U(\bar x)$ denote the minimum $\length(\langle M\rangle)$ 
of a binary encoded Turing machine $M$ such that
$U(\langle M\rangle)$, on empty input,
outputs $\bar x$ and terminates.
Among the properties of this important concept
and the quantity $\Kol_U$, we mention \cite{Vitanyi}:

\begin{fact} \label{f:Discrete}
\begin{enumerate}
\item[a)] 
Its independence, up to additive constants, of the 
universal machine $U$ under consideration.
\item[b)]
The existence and even prevalence of incompressible
instances $\bar x$, that is with $K_U(\bar x)\approx\length(\bar x)$.
\item[c)]
The incomputability (and even Turing-completeness) 
of the function $\bar x\mapsto K_U(\bar x)$;
which is, however, approximable from above.
\item[d)]
Applications in the analysis of algorithms and
the proof of (lower and average) running time bounds.
\end{enumerate}
\end{fact}
We are interested in counterparts to these properties
in the theory of

\subsection{Real Number Computation} \label{s:BSS}
Concerning problems over bits,
the Turing machine is widely agreed
to be the appropriate model of computation:
it has tape cells to hold one bit each,
receives as input and produces as output finite strings over $\{0,1\}$,
can store finitely many of them in its `program code',
and execution basically amounts to the application of 
a finite sequence of Boolean operations. 
A somewhat more convenient model, yet equivalent with respect to
computability, the Random Access Machine (\textsf{RAM})
operates on integers as entities. Both are thus examples of a
model of computation on an algebra: $(\{0,1\},\vee,\wedge,\neg)$
in the first case and $(\IZ,+,-,\times,<)$ in the second.
Among the natural class of such general machines \cite{Zucker},
we are interested in that corresponding to the algebra of
real numbers $(\IR,+,-,\times,\div,<)$: this is known as the
\textsf{real-RAM} and popular for instance in Computational Geometry
\cite{Preparata,deBerg}. In \cite{BSS,BCSS}, it has been re-discovered
and promoted as an idealized abstraction of fixed-precision
floating-point computation.
The latter publication(s) led to the
name ``\BSS model'' which we also adopt in the present work:

\begin{definition} \label{d:BSS}
A \BCSS machine $\IM$ consists of 
\begin{enumerate}
\item[i)] 
An unbounded (input, work, and output) tape
capable of holding a real number in each cell.
\item[ii)]
A reading and a writing head to move independently.
\item[iii)]
A finite set $Q$ of states.
\item[iv)]
A finite, numbered sequence $(c_1,\ldots,c_J)$ of
real constants.
\item[v)]
And a finite control $\delta$ describing,
when in state $q$ and depending on the \emph{sign} of the real $x$
contained in the cell at the reading head's current position,
which of the following actions to take:
\begin{itemize}
\item Copy, add, or multiply $x$ 
  to the real $y$ under the writing head.
\item Subtract $x$ from $y$ or divide $y$ by $x$
 (the latter under the provision that $x\not=0$).
\item Copy some $c_j$ to $y$.
\item Move the reading or writing head one cell
  to the left or to the right.
\item Halt.
\end{itemize}
\end{enumerate}
Let $\IR^*:=\bigcup_{n \in \IN} \IR^n$ denote
the set of finite sequences of real numbers
and $\size(\vec x)=n$ for $\vec x\in\IR^n$.
$\IM$ realizes a partial real function on $\IR^*$
(by abuse of notation also called $\IM:\subseteq\IR^*\to\IR^*$,
$\vec x\mapsto\IM(\vec x)$)
according to the following semantics:
\\
For $\vec x\in\IR^n$,
execution starts with the tape containing $(n,x_1,\ldots,x_n)$.
If $\IM$ eventually terminates \emph{and} the
tape contents is of the form $(m,y_1,\ldots)$
with $m\in\IN$, then $\IM(\vec x):=(y_1,\ldots,y_m)$;
otherwise $\IM(\vec x):=\bot$ (i.e. $\vec x\not\in\dom(\IM)$).

A subset $\IL\subseteq\IR^*$ is called a (real) \emph{language}.
It is (\BCSS) \emph{semi-decidable} if $\IL=\dom(\IM)$
for some \BCSS machine $\IM$. $\IL$ is (\BCSS) \emph{decidable} if
its characteristic function is realized by some $\IM$.
$\IL$ being (\BCSS) \emph{enumerable} means that $\IL=\range(\IM)$
for some total (!) $\IM$.
\end{definition}
The above definition refers to the \BCSS equivalent of a
one-tape two-head Turing machine. It generalizes
to $k$ tapes: as usual without significantly increasing the
power of this model. 
In \cite{BSS,BCSS}, the authors transfer several important concepts 
and results from the classical (i.e. discrete) theory of computation
to the real setting, such as
\begin{itemize}
\item[\textbullet]
  The existence of a universal \BSS machine, capable
  of simulating any given machine and satisfying
  SMN and UTM-like properties.
\item[\textbullet]
  The undecidability of the 
  termination of a given (encoding of another) \BSS machine,
  i.e. of the \emph{real} Halting problem $\IH$.
\item[\textbullet]
  A real language decidable in polynomial time by a
  \emph{non-}deterministic \BSS machine can also be decided
  in exponential time by a \emph{deterministic} one:
\begin{equation} \label{e:oneMillion}
  \textsf{P}_{\IR} \;\subseteq\; \textsf{NP}_{\IR}\;\subseteq\;\textsf{EXP}_{\IR}
  \enspace .
\end{equation}
\item[\textbullet]
  There exist decision problems \emph{complete} for
  $\textsf{NP}_{\IR}$; and, relatedly, an important open question asks
  whether and which of the inclusions in Equation~(\ref{e:oneMillion})
  are strict.
\end{itemize}
Here, running times and asymptotics are considered in terms
of the \emph{size} $n$ of the input $\vec x=(x_1,\ldots,x_n)\in\IR^*$:
a natural algebraic counterpart to the (bit-) \emph{length} 
of binary Turing machine inputs $\bar x=(x_1,\ldots,x_n)\in\{0,1\}^*$.

In fact the last two items above have 
spurred the development of a rich theory of computational
complexity over the reals with classes like 
$\textsf{\#P}_{\IR}$ \cite{Meer,CountingII},
$\textsf{PSPACE}_{\IR}$ \cite{CuckerKoiran95,Perifel},
$\textsf{BPP}_{\IR}$ \cite{Karpinski}, or
$\textsf{PCP}_{\IR}$ \cite{MeerPCP}
and their relations to the discrete realm 
\cite{BuergisserHabil,Buergisser00,Fournier,Buergisser07}.
It is in a certain sense quite surprising 
(and usually rather involved to establish)
that this theory of real computation
exhibits so many properties similar to its
classical counterpart, because proofs
of the latter generally do \emph{not} carry over.
For instance, \textsf{Hilbert's Tenth Problem}
(i.e. the question whether a system of polynomial
equations over field $F$ admits a solution in $F$)
is undecidable over $F=\{0,1\}$ \cite{Matiyasevich}
but for $F=\IR$ becomes decidable due to 
\textsf{Quantifier Elimination}.

\subsection{Pure Algebra} \label{s:Algebra}
This section recalls some well-known mathematical notions and facts;
see for instance \cite{Cohn3,Lang}.

\begin{definition} \label{d:Transc}
Let $E\subseteq F$ denote fields.
\begin{enumerate}
\item[a)]
Call $x\in F$ \emph{algebraic over $E$}
if $p(x)=0$ for some non-zero $p\in E[X]$.
Otherwise $x$ is \emph{transcendental} (over $E$).
\item[b)]
We say that $\{x_1,\ldots,x_n\}\subseteq F$ is
\emph{algebraically dependent over $E$}
if $p(x_1,\ldots,x_n)=0$ for some non-zero $p\in E[X_1,\ldots,X_n]$.
\\
A 
set $X\subseteq F$ is \emph{algebraically dependent over $E$}
if some finite subset of it is.
Otherwise $X$ is called algebraically \emph{in}dependent.
\item[c)]
The \emph{transcendence degree} of $X\subseteq F$ (over $E$),
$\trdeg_E(X)$, is the maximum cardinality of a subset $Y$ of $X$
algebraically independent (over $E$).
\item[d)]
A \emph{transcendence basis} of $F$ (over $E$) is a maximal
algebraically independent subset of $F$.
\item[e)]
$F$ is \emph{purely transcendental} over $E$ if $F = E(S)$ for some $S \subseteq F$ that is algebraically independent over $E$.
\end{enumerate}
\end{definition}
\begin{fact} \label{f:Algebra}
\begin{enumerate}
\item[a)]
Let $a_1,\ldots,a_n\in F$ be algebraic over $E$.
Then there exists some $a\in F$, called a \emph{primitive}
element, such that $E(a_1,\ldots,a_n)=E(a)$.
\item[b)]
If $Y\subseteq X$ is algebraically independent over $E$
and $\Card(Y)=\trdeg_E(X)$, then every element of $X$
is algebraic over $E(Y)$. 
\item[c)] Two transcendence bases have equal cardinality.
\item[d)]
For a chain $E\subseteq F\subseteq G$ of fields,
it holds $\trdeg_E(G)=\trdeg_E(F)+\trdeg_F(G)$.
\item[e)] In $\IR$, $e$ and $\pi$ are transcendental over $\IQ$.
\item[f)] Let $a_1,\ldots,a_n$ be algebraic yet linearly independent over $\IQ$.
Then $e^{a_1},\ldots,e^{a_n}$ are algebraically independent over $\IQ$
\end{enumerate}
\end{fact}
Claim~f) is the \textsf{Lindemann-Weierstra\ss{} Theorem},
cf. e.g. \mycite{Theorem~1.4}{Baker}.

\subsection{Real Kolmogorov Complexity}
The similarities between the discrete theory of Turing computation
and the real one of \BSS machines (Section~\ref{s:BSS}) have led
\person{Monta\~{n}a} and \person{Pardo}
to introduce and study in \cite{Montana}
the following real counterpart to 
classical Kolmogorov complexity:

\begin{definition} \label{d:Kolmogorov}
For a universal \BSS machine $\IU$
and for $\vec x\in\IR^*$ let $\KOL_{\IU}(\vec x)\in\IN$
denote the minimum $\size(\vec p)$, $\vec p\in\IR^*$, such that
$\IU(\vec p)$, on empty input, outputs $\vec x$ and terminates.
\end{definition}
Based on Item~a) in Section~\ref{s:BSS},
they conclude in \mycite{Theorem~2}{Montana}
that Fact~\ref{f:Discrete}a) 
carries over from the discrete to the real setting:

\begin{observation} \label{o:Invar}
For another universal machine $\IU'$, 
$\KOL_{\IU}(\vec x)$ differs from $\KOL_{\IU'}(\vec x)$
only by an additive constant independent of $\vec x$.
\end{observation}
Moreover for the special case of the constant-free universal \BSS machine
$\IU_0$ introduced in \mycite{Section~8}{BSS}, 
\cite[\textsc{Theorems}~3 and 6]{Montana} establish the
real Kolmogorov complexity to be bounded from
below, and up to an additive constant from above, 
by the transcendence degree:

\begin{fact}  \label{f:Trdeg}
There exists some $c\in\IZ$ such that,
for any $\vec x\in\IR^*$, it holds
\begin{equation} \label{e:Trdeg}
\trdeg_{\IQ}(\vec x) \;\leq\; \KOL_{\IU_0}(\vec x)\;\leq\;\trdeg_{\IQ}(\vec x)\,+\,c 
\enspace . \end{equation}
\end{fact}
As an application, \mycite{Corollary~4}{Montana}
presents an alternative proof to a known lower bound
in the algebraic complexity theory of polynomials,
thus exemplifying the real incompressibility method
as a natural counterpart to Fact~\ref{f:Discrete}d).
We will give another application in Observation~\ref{o:Pairing}.

A further consequence of Fact~\ref{f:Trdeg}:
Since a `random' $n$-element real vector has 
transcendence degree equal to $n$,
incompressible strings are prevalent---a counterpart
to Fact~\ref{f:Discrete}b), 
however based on entirely different arguments;
see also Corollary~\ref{c:Prevalent} below.
Moreover, as opposed to the discrete case,
one can explicitly write down such instances,
compare \mycite{Theorem~8}{Montana}
and Example~\ref{x:Transc}a) below.

\subsection{Overview}
We focus on a natural variant of the universal machine $\IU_0$
which leads to particularly compact \BSS programs:
all discrete code information (i.e. anything except for 
the real constants) is encoded into the first real number.
For this G\"{o}delization,
we extend the results in \cite{Montana} in five directions.

First, Fact~\ref{f:Trdeg} can be improved
in that the constant $c$ may be chosen as $1$;
and we show that this is generally best possible.
Second, in Section~\ref{s:Segre}, we consider
the mathematical question in which cases the first inequality 
of Equation~(\ref{e:Trdeg}) is tight and in which
cases the second one; the answer turns out to be
related to deep issues in algebraic geometry.
Then we investigate the computational properties 
of the real Kolmogorov complexity function $\KOL$:
The classical incomputability argument, being based
on exhaustively searching for an incompressible string,
does not carry over to this continuous setting.
Our third contribution features an entirely 
different proof establishing,
as a partial analogue to Fact~\ref{f:Discrete}c), 
the \BSS incomputability of $\KOL$
(Section~\ref{s:Uncomputable}).
Fourth, we show that $\KOL$ can
(as in the discrete case but again by different arguments)
be approximated from above.
And finally in Section~\ref{s:Incomplete}, 
$\KOL$ is proven \emph{not} \BSS-\emph{complete}.

\section{Compact \BSS G\"{o}delization} \label{s:Compact}
While Observation~\ref{o:Invar} asserts a certain invariance
of the Kolmogorov complexity of all strings,
a fixed $\vec x$'s complexity on the other hand may change
dramatically when proceeding from $\IU$ to $\IU'$: simply by
constructing $\IU'$ to give this particular $\vec x$
a special short code treated separately.
Nevertheless, and as opposed to the classical case,
we will now introduce a particular class
of universal real machines $\IU$ and show them 
to give rise to relatively `minimal' $\KOL_{\IU}$:
\begin{definition} \label{d:UTM}
Fix a finite choice $\vec z:=(z_1,\ldots,z_D)$ of reals
and let $\IU_{\vec z}$ 
denote a universal \BSS machine with constants $z_1,\ldots,z_D$
to simulate, upon input of `program'
$\langle\IM\rangle_{\vec z}$ and of $\vec x\in\IR^*$,
$\IM$ on $\vec x$. 
(The empty program produces no output
and terminates precisely on the empty input.)
Here, $\langle\IM\rangle_{\vec z}$ is defined as follows:

Consider a \BCSS-computable integer/real pairing function
$\langle\,\cdot\,,\,\cdot\,\rangle:\IN\times\IR\to\IR$ 
with computable inverse; for instance something like
\[ (n,x)\;\mapsto\;\sign(x)\cdot
\big(2^n\cdot(2\lfloor |x|\rfloor+1)\,+\,(|x|-\lfloor |x|\rfloor)\big)
\enspace . \]
Encode some machine $\IM$,
with constants $c_1,\ldots,c_J,z_1,\ldots,z_D$ and control $\delta$
according to Definition~\ref{d:BSS},
as $\langle\IM\rangle_{\vec z}:=(\langle\delta,c_1\rangle,c_2,\ldots,c_J)$.

Finally
abbreviate $\KOL_{\vec z}:=\KOL_{\IU_{\vec z}}$ and $\KOL_0:=\KOL_{()}$.
\end{definition}
Here we have exploited that the \emph{control} of $\IM$
contains no real constants by itself 
but just \emph{references} to them: to $c_j$ by virtue of
an index $j\in\{1,\ldots,J\}$; or to $z_d$ provided by its `host'
machine $\IU_{\vec z}$ by virtue of an index
$d\in\{1,\ldots,D\}$. That $\delta$ 
thus being a purely discrete object permits
to combine it with one other real, thus saving 1 element in size.
\begin{remark} \label{r:Prefix}
More precisely, \emph{any} finite information (like, e.g. the number
$J$ of real constants following or the length of the input $\vec x$
to simulate $\IM$ on) can be incorporated in
this way without increasing the size of the encoding.
This simplifies several putative pitfalls from classical Kolmogorov Complexity
like \mycite{Example~2.1.4}{Vitanyi} 
\[
\KOL_{\vec z}(\vec x,\vec y)\;\leq\; \KOL_{\vec z}(\vec x)+\KOL_{\vec z}(\vec y)
\]
and, for instance, lifts the
need for a real counterpart to classical 
\emph{prefix} complexity \mycite{Section~3}{Vitanyi}.
\end{remark}
Also note that a fully real/real pairing function cannot be \BCSS computable:
For instance it follows from the \emph{invariance of domain} principle
in Algebraic Topology that a \BCSS computable function from $\IR\times\IR$
to $\IR$ cannot be injective. Alternatively, Observation~\ref{o:Pairing}
below shows that a \BCSS-computable function from $\IR$ to $\IR\times\IR$
cannot be surjective: with a simple proof based on real Kolmogorov Complexity Theory!

\subsection{Real Kolmogorov Complexity and Transcendence Degree} \label{s:Transc}
Intuitively, the encoding introduced in Definition~\ref{d:UTM} is as `compact' as
possible. Indeed, we have the following

\begin{observation}
For any universal real machine $\IU'$ with constants $\subseteq\{z_1,\ldots,z_D\}$,
it holds $\KOL_{(z_1,\ldots,z_D)}\leq\KOL_{\IU'}$.
\end{observation}
\begin{proof}
Since $\IU_{\vec z}$ already contains all real constants of $\IU'$,
$\langle\IU'\rangle_{\vec z}$ is purely discrete; now apply 
Remark~\ref{r:Prefix}.
\qed\end{proof}
Since we are aiming for bounds on \BCSS Kolmogorov
Complexity that are as tight as possibly, it turns out beneficial
to refine Definition~\ref{d:Kolmogorov}
to distinguish between the following closely
related quantities corresponding to enumerability,
decidability, and semi-decidability:

\begin{definition} \label{d:Kolmogorov2}
\begin{enumerate}
\item[a)]
For $\bar x\in\{0,1\}^*$ let $\KolO_U(\bar x)$ 
denote the minimum $\length(\bar p)$, $p\in\{0,1\}^*$,
such that
$U(\bar p)$, on empty input, \emph{outputs} $\bar x$ and terminates.
\item[b)]
$\KolS_U(\bar x)$ and $\KolD_U(\bar x)$ are defined 
similarly by the condition that $U(\bar p)$
\emph{semi-/decides} the single-word language $\{\vec x\}$.
\item[c)]
For $\vec x\in\IR^*$ let $\KOLO_{\IU}(\vec x)$
denote the minimum $\size(\vec p)$, $\vec p\in\IR^*$, such that
$\IU(\vec p)$, on empty input, \emph{outputs} $\vec x$ and terminates.
\item[d)]
$\KOLS_{\IU}(\vec x)$ and $\KOLD_{\IU}(\vec x)$ are defined similarly
by the condition that $\IU(\vec p)$
\emph{semi-/decides} the single-word language $\{\bar x\}$.
\end{enumerate}
\end{definition}
One usually focuses on $\KolO$ (and we on $\KOLO$).
Indeed, $\KolO_U$, $\KolD_U$, and $\KolS_U$ 
differ at most by an additive constant
independent of $\bar x$: a machine $M$ 
outputting $\bar x$ can be turned (with a fixed
increase in complexity) into one 
which, given $\bar y$, simulates $M$ and compares
its output to the input in order to semi-/decide $\{\bar x\}$; 
conversely, $M$ semi-deciding $\{\bar x\}$ may be
used by $M'$ generating \emph{all} binary strings $\bar y$
to output the one that $M$ terminates on.
In the \BSS realm, the inequality
``$\KOLS_{\IU}(\vec x)\leq\KOLD_{\IU}(\vec x)\leq\KOLO_{\IU}(\vec x)+\calO(1)$''
can be proven similarly; whereas
``$\KOLO_{\IU}(\vec x)\leq\KOLS_{\IU}(\vec x)+\calO(1)$''
requires some more work, because one cannot
generate \emph{all} real strings.
In fact, it is a consequence of Observation~\ref{o:Invar}
and the following, already announced

\begin{theorem} \label{t:Transc}
For every $\vec x\in\IR^+$ and $\vec z\in\IR^*$ it holds
\begin{enumerate}
\itemsep5pt
\item[a)] $\displaystyle 
\KOLS_{\vec z}(\vec x) = \KOLD_{\vec z}(\vec x) = \max\{1,\trdeg_{\IQ(\vec z)}(\vec x)\}$.
\item[b)] $\displaystyle
\max\{1,\trdeg_{\IQ(\vec z)}(\vec x)\} \leq\KOLO_{\vec z}(\vec x)\leq\trdeg_{\IQ(\vec z)}(\vec x)+1$;
\item[c)] If $\IQ(\vec z,\vec x)$ is purely transcendental over $\IQ(\vec z)$,
  then $\KOLO_{\vec z}(\vec x)=\trdeg_{\IQ(\vec z)}(\vec x)$.
\end{enumerate}
\end{theorem}
Section~\ref{s:Proof} contains the proof of this theorem.
\begin{corollary} \label{c:Prevalent}
Incompressible strings exist;
they are in fact prevalent.
\end{corollary}
\begin{proof}
For fixed $z_1,\ldots,z_D,x_1,\ldots,x_n\in\IR$,
the set 
$\{x\in\IR: x\text{ algebraic over }\IQ(\vec z,\vec x)\}$
is countable. Therefore, guessing $x_1,\ldots,x_n\in[0,1]$ inductively
independently
uniformly at random yields with certainty $\trdeg_{\IQ(\vec z)}(\vec x)=n$.
\qed\end{proof}
\begin{example} \label{x:Transc}
\begin{enumerate}
\item[a)]
$\KOLO_0(e^{\sqrt{2}},e^{\sqrt{3}},e^{\sqrt{5}},e^{\sqrt{7}},e^{\sqrt{11}},\ldots,e^{\sqrt{p_n}})=n$,
where $p_n\in\IN$ denotes the $n$-th prime number.
\item[b)] 
For $t\in\IR$, it holds
$\KOLO_0(t,\sqrt{2})=1$ 
in case $t$ is algebraic
and $\KOLO_0(t,\sqrt{2})=2$ 
if $t$ is transcendental.
\end{enumerate}
\end{example}
\begin{proof}
Indeed $\sqrt{2},\sqrt{3},\ldots,\sqrt{p_n}$ are square roots
of distinct square-free numbers and therefore \cite{Rickard}
linearly independent over $\IQ$; from which it follows by
Fact~\ref{f:Algebra}f) that their
exponentials are algebraically independent over $\IQ$.
Now apply Theorem~\ref{t:Transc}c).

The first part of Claim~b) follows immediately
from Theorem~\ref{t:Transc}b); similarly for
the inequality ``$\leq2$'' of the second part.
The reverse inequality is a consequence
of Proposition~\ref{p:Segre}a) below since
$\sqrt{2}\not\in\IQ(t)$ for $t$ transcendental.
Indeed the presumption $\sqrt{2}=p(t)/q(t)$
with polynomials $p,q\in\IQ[T]$ would imply
$p^2(t)=2q^2(t)$, hence $p^2-2q^2$ vanishes
identically: in contradiction to the 
(classical proof of the)
irrationality of $\sqrt{2}$.
\qed\end{proof}
\subsection{Proof of Theorem~\ref{t:Transc}} \label{s:Proof}
\begin{enumerate}
\item[a)]
A machine deciding $\IL$ is easily turned into one \emph{semi-}deciding $\IL$
without introducing any further constant: this shows $\KOLS\leq\KOLD$.

In \cite{Michaux} it has been shown that a language $\IL$ semi-decided by
some \BCSS machine $\IM$ is a countable union of sets basic semi-algebraic 
(i.e. solutions of a system of polynomial in-/equalities) over the
rational field extension $\IQ(y_1,\ldots,y_N)$ generated by the 
real constants $y_1,\ldots,y_N$ of $\IM$;
see also \mycite{Theorem~2.4}{Cucker}. 
Since in our case $\IL=\{\vec x\}$
is a singleton, it must even be basic semi-algebraic. In fact, semi-algebraic
sets being closed under projection \mycite{Section~2.4}{Basu},
each single component $x_1,\ldots,x_n$ is a solution of some polynomial 
in-/equalities over $\IQ(y_1,\ldots,y_N)$. It cannot be inequalities
only, otherwise the solution would be open. Thus $x_1,\ldots,x_n$
are all algebraic over $\IQ(y_1,\ldots,y_N)$.
Applied to the \BCSS machine $\IU_{\vec z}(\vec p)$ with constants
$\{y_1,\ldots,y_N\}=\{\vec z,\vec p\}$ shows that
$\{\vec x\}$ is algebraic over $\IQ(\vec z)(\vec p)$.
Therefore, according to Fact~\ref{f:Algebra},
$\trdeg_{\IQ(\vec z)}(\vec x)\leq\trdeg_{\IQ(\vec z)}(\vec p)\leq\size(\vec p)$
shows $\KOLS_{\vec z}(\vec x)\geq\trdeg_{\IQ(\vec z)}(\vec x)$.
$\KOLS_{\vec z}(\vec x)\geq1$ holds because $\vec x\not=()$ 
requires some coding.

Finally to see $\KOLD_{\vec z}(\vec x)\leq\max\{1,\trdeg_{\IQ(\vec z)}(\vec x)\}$,
first consider the case $\trdeg_{\IQ(\vec z)}(\vec x)=0$.
By Fact~\ref{f:Algebra}b), $x_1,\ldots,x_n$ are all algebraic over $\IQ(\vec z)$.
For each $i=1,\ldots,n$
let $0\not=p_i(\vec z,X)\in\IQ(\vec z)[X]$ denote some polynomial
having $x_i$ as unique root within the interval $(a_i,b_i)$, $a_i,b_i\in\IQ$.
These finitely many rationals $a_i,b_i$ constitute discrete information
only; and so do the coefficients of $p_i$ described in terms of
rational functions over $\vec z$. Since $\vec z$ itself is provided
by the universal host machine $\IU_{\vec z}$, the remaining data
about $p_1,\ldots,p_n$ can be combined into one number
which admits an effective evaluation of $y_i\mapsto p_i(y_i)$
and tests ``$p_i(y_i)=0$, $a_i<y_i<b_i$'' to decide whether
a given input $\vec y$ belongs to $\{\vec x\}$:
$\KOLD_{\vec z}(\vec x)\leq 1$.

In remaining case $d:=\trdeg_{\IQ(\vec z)}(\vec x)>0$,
let $\{p_1,\ldots,p_d\}$ denote some transcendence 
basis of $\IQ(\vec z,\vec x)$ over $\IQ(\vec z)$.
Again by virtue of Fact~\ref{f:Algebra}, all $x_i$ are
algebraic over $\IQ(\vec z,\vec p)$ and describable by
rational bounds and polynomials $p_i(\vec z,\vec p,X)\in\IQ(\vec z,\vec p)[X]$.
By virtue of Remark~\ref{r:Prefix}, this data can be combined
with the $d$ reals $p_1,\ldots,p_d$ to show $\KOLD_{\vec z}(\vec x)\leq d$.
\item[b1)]
The first inequality of b) follows from a) by observing
$\KOLD_{\vec z}\leq\KOLO_{\vec z}$: a machine to output
$\vec x$ can be transformed into one deciding $\{\vec x\}$
incurring only discrete additional cost; now apply Remark~\ref{r:Prefix}.
\item[c)]
Let $p_1,\ldots,p_d$ denote a transcendence
basis of $\IQ(\vec z,\vec x)$ over $\IQ(\vec z)$.
By prerequisite, $x_1,\ldots,x_n$ 
are not only algebraic over (Fact~\ref{f:Algebra}b),
but even belong to, $\IQ(\vec z,\vec p)$.
They can thus be described and computed using,
in addition to $\vec z$ and $\vec p$, only
discrete information. In view of Remark~\ref{r:Prefix},
this shows $\KOLO_{\vec z}(\vec x)\leq\trdeg_{\IQ(\vec z)}(\vec x)$.
\item[b2)]
For the second inequality of b), we proceed similarly to the proof
of Claim~c), however taking into account that now
$x_1,\ldots,x_n$ need not belong to, but are only algebraic over,
$\IQ(\vec z,\vec p)$. On the other hand, by Fact~\ref{f:Algebra}a),
there exists some primitive element
$a\in\IR$ such that $x_1,\ldots,x_n\in\IQ(\vec z,\vec p,a)$.
Now $\vec x$ can be described and computed as above,
using $\vec z$, $\vec p$, \emph{and} $a$.
\qed
\end{enumerate}

\subsection{Non-Purely Transcendental Extensions} \label{s:Segre}
Unless $\vec x$ is purely transcendental, 
Theorem~\ref{t:Transc}b) leaves a gap of 1 between
lower and upper bound. This turns out very difficult
to close and leads to deep questions in algebraic geometry:

\begin{proposition} \label{p:Segre}
\begin{enumerate}
\item[a)]
Let $t\in\IR$ be transcendental over $\IQ(\vec z)$
and $a\not\in\IQ(\vec z,t)$ algebraic over $\IQ(\vec z,t)$.
Then $\KOLO_{\vec z}(t,a)=2>1=\trdeg_{\IQ(\vec z)}(t,a)$.
\item[b)]
To any $s,t\in\IR$ algebraically independent over $\IQ$
there exist $x,y,a\in\IR$ such that 
$s,t,a\in\IQ(x,y)$ and $a\not\in\IQ(s,t)$.
\\
In particular, it holds $\KOLO_0(s,t,a)=2=\trdeg_{\IQ}(s,t,a)$
\emph{although} $a$ is not algebraic over $\IQ(s,t)$.
\end{enumerate}
\end{proposition}
The latter shows that 
there is no ``only if'' 
in Theorem~\ref{t:Transc}c).
\begin{proof}[Proposition~\ref{p:Segre}]
\begin{enumerate}
\item[a)]
Suppose toward contradiction that
some \BSS machine $\IM$ with one real constants $\vec z,x$
can output $t,a$. By induction on the number of steps performed
by $\IM$, it is easy to see that any intermediate result and
in particular its output constitutes a rational function 
of $\vec z,x$, that is, belongs to $\IQ(\vec z,x)$. Since $t\in\IQ(\vec z,x)$
is transcendental over $\IQ(\vec z)$, so must be $x$ itself.
\textsf{L\"{u}roth's Theorem} asserts every 
subfield between $\IQ(\vec z)$ and its simple transcendental extension $\IQ(\vec z,x)$
to be simple again; cf. e.g. \mycite{Theorem~5.2.4}{Cohn3}.
However $\IQ(\vec z,t,a)$ by prerequisite
is not simple over $\IQ(\vec z)$: a contradiction.
\item[b)]
L\"{u}roth's Theorem has been extended by \person{Castelnuovo}
to the case of transcendence degree 2---however over algebraically
\emph{closed} fields. It is now known to fail from transcendence
degree 3 on, and also for 2 over an algebraically \emph{non-}closed
field. See for instance to \mycite{Remarks~6.6.2}{Gille}
for a historical account of these results.

In particular for the field $\IQ$, we refer to a classical 
counter-example \cite{Segre} due to \person{Beniamino Segre}
showing the $\IQ$-variety $V$ defined by the cubic 
$b^3+3a^3+5s^3+7t^3$
on the $\IQ$-sphere $\calS^3=\{(a,b,s,t)\in\IQ^4:a^2+b^2+s^2+t^2=q^2\}$,
$q\in\IQ$, to be unirational but not rational.
In other words (cmp. Lemma~\ref{l:Indep}a below): 
For arbitrary $s,t$ transcendental over $\IQ$
and sufficiently large $q$,
a (thus real) solution $a$ to
$q^2-a^2-s^2-t^2 = (3a^3+5s^3+7t^3)^2$
is algebraic over (but not contained in) 
$\IQ(s,t)$; whereas unirationality of $V$ means
that $\IQ(s,t,a)$ be in turn contained
in some purely transcendental extension $\IQ(x,y)$.
A \BSS machine storing $x,y$
can therefore output $s,t,a$ as rational functions thereof,
showing $\KOLO_0(s,t,a)\leq2$.
\qed\end{enumerate}
\end{proof}
\section{Incomputability} \label{s:Uncomputable}
A folklore property of classical Kolmogorov Complexity is its
incomputability: No Turing machine can evaluate
the function $\{0,1\}^*\ni\bar x\mapsto\Kol(\bar x)$. This follows
from a formal argument related to the \textsf{Richard-Berry Paradox}
which involves a contradiction arising from searching
for some $\bar x\in\{0,1\}^*$ of minimum length $n$
such that $\Kol(\bar x)$ exceeds a given bound;
cf. e.g. \mycite{Theorem~5.5}{Moret}.

\begin{remark} \label{r:Uncomputable}
Over the reals, as opposed to $\{0,1\}^n$, 
$\IR^n$ is too `large' to be searched.
As a consequence, concerning the simulation of 
a nondeterministic \BCSS machine by deterministic one,
based on Tarski's Quantifier Elimination as in 
\mycite{Section~2.5.1}{Basu}
the \emph{existence} of a successful real guess can
be decided, but a \emph{witness} can in general not be found.
More precisely, a \BCSS machine with constants $c_1,\ldots,c_J$
is limited to generate numbers in $\IQ(c_1,\ldots,c_J)$
(compare the proof of Proposition~\ref{p:Segre}a)
and thus cannot \emph{output}, even with the help of
oracle access to $\KOLO$, any real vector of
Kolmogorov Complexity exceeding $J$ in order to raise
a contradiction to the presumed computability of $\KOLO$.

Similarly, the classical proof does not carry over
to show the incomputability of the \emph{decision} version
$\KOLD$, either: Given $\vec x$ as \emph{input} one can, 
relative to $\KOLD$, detect (and terminate, provided) that 
$\vec x$ has sufficiently high Kolmogorov Complexity;
however this approach accepts a large, not a one-element
real language.
\qed\end{remark}
Nevertheless we succeed in establishing

\begin{theorem} \label{t:Uncomputable}
For each $\vec z\in\IR^*$,
both $\KOLO_{\vec z}$ and $\KOLD_{\vec z}$
are \BCSS--\emph{in}computable,
even when restricted to $\IR^2$.
\end{theorem}
The proof is based on Claim~c) of the following
\begin{lemma} \label{l:Undecidable}
\begin{enumerate}
\item[a)] The set $\IT\subseteq\IR$ of 
  transcendental reals (over $\IQ$) is not \BCSS semi-decidable.
\item[b)] $\IT$ is not even semi-decidable
  relative to oracle $\IQ$.
\item[c)]
For $\vec y,\vec z\in\IR^*$, the real language
~$\IT_{\vec z}:=\{x\in\IR: x\text{ transcendental over }\IQ(\vec z)\}$~
is not \BCSS semi-decidable relative to oracle $\IQ(\vec y)$.
\item[d)]
For $\vec z\in\IR^*$, the real language
$\IR\setminus\IT_{\vec z}=\{x\in\IR:x\text{ algebraic over }\IQ(\vec z)\}$
is \BCSS semi-decidable.
\end{enumerate}
\end{lemma}
Claim~a) is folklore. Its extension b)
has been established as \mycite{Theorem~4}{realpost}
and generalizes straight-forwardly to yield Claim~c).
Here we implicitly refer to the concept of 
\BCSS \emph{oracle} machines $\IM^{\IO}$ whose
transition function $\delta$ may, in addition 
to Definition~\ref{d:BSS}v), enter a \textsf{query}
state corresponding to the question
whether the contents of the dedicated query tape 
belongs to $\IO\subseteq\IR^*$, and proceed according to
the (Boolean) answer.

Regarding Claim~d) it suffices to enumerate all 
non-zero $p\in\IQ(\vec z)[X]$ and test ``$p(x)=0$''.

\begin{proof}[Theorem~\ref{t:Uncomputable}]
Concerning $\KOLD_{\vec z}$, fix some $s\in\IR$ transcendental
over $\IQ(\vec z)$. Then, according to
Theorem~\ref{t:Transc}a), $\KOLD_{\vec z}(s,t)=2$ if
$t\in\IT_{\vec z,s}$, and $\KOLD_{\vec z}(s,t)=1$ otherwise;
that is \BCSS-computability of $\KOLD_{\vec z}(s,\cdot)$
contradicts Lemma~\ref{l:Undecidable}c).

Similarly, according to Example~\ref{x:Transc}b),
$\KOLO_{\vec z}(t,\sqrt{2})=2$ if
$t\in\IT_{\vec z}$, and $\KOLO_{\vec z}(t,\sqrt{2})=1$
otherwise.
\qed\end{proof}

\subsection{Approximability}
Although the function $\bar x\mapsto\Kol(\bar x)$ is not Turing-computable,
it can be approximated \mycite{Theorem~2.3.3}{Vitanyi}: 
from above, in the point-wise limit without error bounds.

\begin{fact} \label{f:Approx}
The set $\{(\bar x,k):\Kol(\bar x)\leq k\}\subseteq\{0,1\}^*\times\IN$ 
is semi-decidable.
\end{fact}
In particular
$\Kol$ becomes computable given oracle access to the Halting problem $H$.%

\begin{fact}[Shoenfield's Limit Lemma] \label{f:Shoenfield}
A function $f:\subseteq\{0,1\}^*\to\IN$ is computable
\emph{relative} to $H$ ~iff~ $f(\bar x)=\lim_{m\to\infty} g(\bar x,m)$
for some \emph{ordinarily} computable $g:\dom(f)\times\IN\to\IN$.
\end{fact}
See for instance \mycite{\textsection III.3.3}{Soare}\ldots

\begin{remark} \label{r:Shoenfield}
Concerning a real counterpart of Fact~\ref{f:Shoenfield},
only the domain but not the range extends from discrete to $\IR$:
\begin{enumerate}
\item[a)]
A function $f:\IR^*\to\IN$ is \BCSS computable relative to 
the \emph{real} Halting Problem 
\[ \IH \;=\; \big\{\langle\IM\rangle:\IM\text{ terminates on input }()\big\} \]
~iff~ $f(\vec x)=\lim_{m\to\infty} g(\vec x,m)$
for some \BCSS computable $g:\dom(f)\times\IN\to\IN$.
\item[b)]
The function $\exp:\IR\ni x\mapsto e^x\in\IR$ is the point-wise
limit of \BCSS-computable $g(x,m):=\sum_{n=0}^{m} x^n/n!\in\IR$; 
$\exp$ is, however,
not \BCSS-computable relative to any oracle $\IO\subseteq\IR^*$.
\end{enumerate}
\noindent
Computing real limits is the distinct feature of
so-called \emph{Analytic} Machines \cite{Hotz}.
\end{remark}
\begin{proof} 
\begin{enumerate}
\item[a1)]
Since $g(\vec x,\cdot)$ has discrete range,
the sequence $\big(g(\vec x,m)\big)_{_m}$ must eventually
stabilize to its limit $f(\vec x)$.
Now the real UTM and SMN theorems make it easy to 
construct from $\vec x\in\IR^*$ and $M\in\IN$ a
\BCSS machine $\IM$ which terminates iff
$\big(g(\vec x,m)\big)_{_{m\geq M}}$ is not constant.
Repeatedly querying $\IH$ thus allows to determine
$\lim_{m\to\infty} g(\vec x,m)=f(\vec x)$.
\item[a2)]
Let $f$ be computable relative to $\IH$ by
\BCSS oracle machine $\IM^{\IH}$.
Given $\vec x\in\dom(f)$, $\IM^{\IH}$ thus makes
a finite number (say $N$) of steps and oracle queries;
let $\vec u_1,\ldots,\vec u_N\in\IH$ denote those answered
positively and $\vec v_1,\ldots,\vec v_N\not\in\IH$ those
answered negatively.
Now define $g(\vec x,m)$ as the output of the following
computation: Simulate $\IM$ for at most $m$ steps and,
for each oracle query ``$\vec w\in\IH$?'', perform the
first $m$ steps of a semi-decision procedure:
if it succeeds, answer positively, otherwise negatively.
\\
Now although the latter answer may in general be wrong,
the finitely many queries $\vec u_1,\ldots,u_N\in\IH$
admit a common $M$ beyond which all are reported
correctly; and so are the negative ones $\vec v_j\not\in\IH$
anyway. Hence for $m\geq M,N$, $g(\vec x,m)=f(\vec x)$.
\item[b)]
The proof of Proposition~\ref{p:Segre}a) 
has already exploited that all intermediate
results (and in particular the output $y$), computed by
a \BSS machine with constants $\vec c$ upon
input $\vec x$, belong to $\IQ(\vec c,\vec x)$
and in particular satisfy
$\trdeg_{\IQ}(\vec y)\leq\trdeg_{\IQ}(\vec c,\vec x)
\leq\size(\vec c)+\trdeg_{\IQ(\vec c)}(\vec x)$
according to Fact~\ref{f:Algebra}d);
whereas, for $(x_n):=(\sqrt{2},\sqrt{3},\sqrt{5},\sqrt{7},\sqrt{11},\ldots)$ 
denoting the sequence of square roots of prime integers, 
the corresponding values $y_n:=\exp(x_n)$ 
have according to Fact~\ref{f:Algebra}f)
transcendence degree unbounded
compared to $\trdeg(x_n)=0$.
\qed\end{enumerate}\end{proof}
We now establish a real version of Fact~\ref{f:Approx}.
\begin{proposition} \label{p:Approx}
Fix $\vec z\in\IR^*$.
\begin{enumerate}
\item[a)]
The \emph{real Kolmogorov set}
$\ISD_{\vec z}:=\{(\vec x,k):\KOLD_{\vec z}(\vec x)\leq k\}\subseteq\IR^*\times\IN$ 
is \BCSS semi-decidable.
\item[b)]
$\KOLD_{\vec z}:\IR^*\to\IN$ is \BCSS-computable
\emph{relative} to $\IH$.
\end{enumerate}
\end{proposition}
By virtue of Remark~\ref{r:Shoenfield}a),
Claim~b) follows from a); which in turn is based on 
Lemma~\ref{l:Undecidable}d) in combination with 
Part~b) of the following

\begin{lemma}
\begin{enumerate}
\item[a)]
Let $U$ denote a vector space and 
$V=\lspan(\vecbf y_1,\ldots,\vecbf y_n)\subseteq U$ 
the subspace spanned by
$\vecbf y_1,\ldots,\vecbf y_n\in U$. Then
\begin{multline*} \dim(V) \;=\; n - \max\big\{k\:\big|\:
 \exists 1\leq i_1<\ldots<i_k\leq n: \\
\forall j\in\{1,\ldots,n\}\setminus\{i_1,\ldots,i_k\}:
  \vecbf y_j\in\lspan(\vecbf y_{i_1},\ldots,\vecbf y_{i_k})\big\}
\end{multline*}
\item[b)]
Let $F=E(y_1,\ldots,y_n)$ denote a finitely generated field
extension. Then 
\begin{multline*} \trdeg_E(F) \;=\; n - \max\big\{k\:\big|\:
 \exists 1\leq i_1<\ldots<i_k\leq n: \\
\forall j\in\{1,\ldots,n\}\setminus\{i_1,\ldots,i_k\}:
  y_j\text{ algebraic over } E(y_{i_1},\ldots,y_{i_k})\big\}
\end{multline*}
\end{enumerate}
\end{lemma}
Part~a) is of course the rank-nullity theorem from highschool linear algebra
and mentioned only in order to point out the
similarity to b).
\begin{proof}
Any $y_j$ algebraic over $E(y_{i_1},\ldots,y_{i_k})$
cannot be part of a transcendence basis; hence
$\trdeg_E(F)\leq n-k$. Conversely, choosing
$(y_{i_1},\ldots,y_{i_k})$ as a transcendence basis
yields $\trdeg_E(F)\geq n-k$ according to Fact~\ref{f:Algebra}.
\qed\end{proof}

\subsection{(Lack of) Completeness} \label{s:Incomplete}
Classically, undecidable problems are `usually' also 
Turing-complete in the sense of admitting a (Turing-) reduction
to the discrete Halting problem $H$.
This holds in particular for the Kolmogorov Complexity function;
cf. e.g. \mycite{Exercise~2.7.7}{Vitanyi}.
Over the reals on the other hand, 
$\IQ$ has been identified in 
\cite{realpost} as a decision problem 
\BCSS undecidable but \emph{not} complete.
Similarly, 
\BCSS incomputability of $\KOLD$ according to
Theorem~\ref{t:Uncomputable} turns out to 
\emph{not} extend to \BCSS completeness:

\begin{theorem} \label{t:Incomplete}
Fix $\vec z\in\IR^*$.
\begin{enumerate}
\item[a)] Let
\[ \II_{\vec z} \;:=\; \big\{\vec x\in\IR^*: \vec x\text{ algebraically
independent over }\IQ(\vec z)\big\} \enspace . \]
Then $\ISD_{\vec z}$ is decidable relative to $\II_{\vec z}$
and vice versa.
\item[b)]
Let $C\subseteq[0,1]$ denote \textsf{Cantor's Excluded Middle Third},
that is the set of all $x=\sum_{n=1}^\infty t_n3^{-n}$ with $t_n\in\{0,2\}$.
Then $C$'s complement is \BCSS semi-decidable
\item[c)]
but $C$ itself is not semi-decidable even relative to $\II_{\vec z}$.
\item[d)]
$\IH$ is not decidable relative to $\ISD_{\vec z}$ 
or to $\KOLD_{\vec z}$.
\end{enumerate}
\end{theorem}
\begin{lemma} \label{l:Cantor}
Fix $\vec w\in\IR^*$.
\begin{enumerate}
\item[a)]
To $x\in C$ and $\epsilon>0$, there exists
$y\in\IT_{\vec w}\setminus C$ with $|x-y|\leq\epsilon$.
\item[b)] The set
$C\cap\IT_{\vec w}$ is uncountable and perfect
(i.e. to $\epsilon>0$ and $x\in C\cap\IT_{\vec w}$
there exists $y\in C\cap\IT_{\vec w}$ with $0<|x-y|\leq\epsilon$).
\end{enumerate}
\end{lemma}
\begin{proof}
Notice that $\IR\setminus\IT_{\vec w}$ is only countable.
\begin{enumerate}
\item[a)]
Let $x=\sum_{n=1}^\infty t_n3^{-n}$ with $t_n\in\{0,2\}$
and $\epsilon=3^{-N}$. The open interval
$I_{x,N}:=\sum_{n=1}^{N-1} t_n3^{-n}+3^{-N}\cdot(\tfrac{1}{3},\tfrac{2}{3})$
is disjoint from $C$ and uncountable;
hence so is $I_{x,N}\setminus(\IR\setminus\IT_{\vec w})$.
From the latter, choose any $y$: done.
\item[b)]
Since $C$ is uncountable, so must be 
$C\setminus (\IR\setminus\IT_{\vec w})$.

Let $x=\sum_{n=1}^\infty s_n3^{-n}$ with $s_n\in\{0,2\}$
and $\epsilon=3^{-N}$. 
Already knowing that $C\cap\IT_{\vec w}$ is infinite,
we conclude that there exists
some $y'=\sum_{n=1}^\infty t_n3^{-n}\in C\cap\IT_{\vec w}$
distinct from $x$ with $t_n\in\{0,2\}$.
Now let $y:=\sum_{n=1}^N s_n3^{-n} + \sum_{n=N+1}^\infty t_{n-N}3^{-n}$:
It satisfies $|x-y|\leq\epsilon$,
belongs to $C$ (having ternary expansion consisting only
of 0s and 2s) and to $\IT_{\vec w}$
(since it differs from $y\in\IT_{\vec w}$ by a rational 
scaling and rational offset).
\qed\end{enumerate}
\end{proof}

\begin{proof}[Theorem~\ref{t:Incomplete}]
\begin{enumerate}
\item[d)] Since $C$ is decidable relative to $\IH$ (b),
$\IH$ cannot be decidable relative to $\II_{\vec z}$ 
(by b) or (by a) to $\ISD_{\vec z}$ or to $\KOLD_{\vec z}$.
\item[a)]
By Theorem~\ref{t:Transc}a) for $\vec x\in\IR^n$, 
$\vec x\in\II_{\vec z}\Leftrightarrow(\vec x,n)\in\ISD_{\vec z}$.
Conversely, $\KOLD_{\vec z}(\vec x)$ can be computed
(and ``$(\vec x,k)\in\ISD_{\vec z}$'' thus decided) by
finding the maximal $k$ such that there exist integers
$1\leq n_1<\ldots<n_k\leq n$ with $(x_{n_1},\ldots,x_{n_k})\in\II_{\vec z}$.
\item[b)]
$[0,1]\setminus C$ is semi-decidable as 
the union of countably many open intervals 
$\sum_{n=1}^N t_n3^{-n}+3^{-N}\cdot(\tfrac{1}{3},\tfrac{2}{3})$,
$N\in\IN$, $t_1,\ldots,t_N\in\{0,2\}$.
\item[c)]
Suppose machine $\IM$ with constants $c_1,\ldots,c_J$
and supported by oracle $\II_{\vec z}$ semi-decides $C$.
Unrolling its computations on all inputs $x\in\IR$ leads
to an infinite 6-ary tree whose nodes $u$ are labelled
with (vectors of) rational functions $f_u\in\IQ(\vec c,X)$ 
meaning that $\IM$ branches on the sign of $f_u(\vec c,x)$
and depending on whether $\vec f_u(\vec c,x)\in\II_{\vec z}$.
Moreover, by hypothesis, the path in this tree taken by
input $x$ ends in a leaf ~iff~ $x\in C$.

Fix some $x\in C$ transcendent over $\IQ(\vec c,\vec z)$ according to
Lemma~\ref{l:Cantor}b). Then $f_u(\vec c,x)\not=0$ for
all $u$ on the finite path $(u_1,\ldots,u_I)$ taken by $x$.
Therefore the set 
\[ 
\{y\in\IR:\sign f_{u_i}(\vec c,y)=\sign f_{u_i}(\vec c,x), i=1,\ldots,I\}
\] 
is open (and non-empty).
Hence, by Lemma~\ref{l:Cantor}a), there are (plenty of)
$y\in\IT_{(\vec c,\vec z)}\setminus C$ belonging to this set.
Moreover, for any such $y$ it holds
$\vec f_u(\vec c,x)\in\II_{\vec z}\Leftrightarrow
\vec f_u(\vec c,y)\in\II_{\vec z}$ according to Lemma~\ref{l:Indep}a) below.
We conclude that $y$ takes the very same path (i.e. follows
the same computation of $\IM$) as $x$: although $x\in C$
and $y\not\in C$, a contradiction.
\qed\end{enumerate}
\end{proof}

\begin{lemma} \label{l:Indep}
Let $E\subseteq F$ denote infinite fields.
\begin{enumerate}
\item[a)]
Fix $x\in F$ transcendental over $E$ and $p_1,\ldots,p_n\in E[X]$.
Then the vector of `numbers' $\big(p_1(x),\ldots,p_n(x)\big)\in E(x)^n$ is algebraically 
independent over $E$ ~iff~ the vector of `functions'
$(p_1,\ldots,p_n)\in E(X)^n$ is.
\item[b)] Fix $\calX,\calY\subseteq F$, $\calX$ algebraically independent over $E$.
Then $\calX\cup\calY$ is algebraically in-/dependent over $E$ 
~iff~ $Y$ is algebraically in-/dependent over $E(\calX)$.
\item[c)]
Let $p\in E[X_1,\ldots,X_n,Y_1,\ldots,Y_m]$ and
$x_1,\ldots,x_n\in F$ be algebraically independent over $E$.
Then $p$ is irreducible (in $E[X_1,\ldots,X_n,Y_1,\ldots,Y_m]$)
~iff~ 
$p(x_1,\ldots,x_n,\cdots)$ is irreducible in $E(x_1,\ldots,x_n)[Y_1,\ldots,Y_m]$.
\item[d)]
Let $p\in E[X_1,\ldots,X_n,Y_1,\ldots,Y_m,Z]$ be irreducible and
$x_1,\ldots,x_n,y_1,\ldots,y_m\in F$ algebraically independent over $E$
but $y_1,\ldots,y_m,z\in F$ algebraically dependent over $E$
and $p(x_1,\ldots,x_n,y_1,\ldots,y_m,z)=0$.
Then $p$ does not `depend' on $X_1,\ldots,X_n$,
i.e. belongs to $E[Y_1,\ldots,Y_m,Z]$.
\end{enumerate}
\end{lemma}
\begin{proof}
\begin{enumerate}
\item[a)]
If $(p_1,\ldots,p_n)$ are algebraically dependent,
say $q(p_1,\ldots,p_n)=0$ for $0\not=q\in E[X_1,\ldots,X_n]$,
then \emph{a fortiori} $q\big(p_1(x),\ldots,p_n(x)\big)=0$.

Conversely let $q\big(p_1(x),\ldots,p_n(x)\big)=0$ for some non-zero
$q\in E[X_1,\ldots,X_n]$. Then $q(p_1,\ldots,p_n)\in E[X]$
vanishes on $x$. Since $x$ is by hypothesis transcendental over $E$,
this implies $q(p_1,\ldots,p_n)=0$.
\item[b)]
Let $\calY$ be algebraically dependent over $E(\calX)$,
$0=p(y_1,\ldots,y_m)$ for $0\not=p\in E(\calX)[Y_1,\ldots,Y_m]$ where 
\begin{gather*} n\in\IN, \quad 
p=\sum_{\bar\imath} \frac{q_{\bar\imath}(x_1,\ldots,x_n)}{r_{\bar\imath}(x_1,\ldots,x_n)}
\cdot Y^{\bar\imath}, \quad
x_1,\ldots,x_n\in\calX, \quad \\ \text{and }\quad 
q_{\bar\imath},r_{\bar\imath}\in E[X_1,\ldots,X_n],
r_{\bar\imath}(x_1,\ldots,x_n)\not=0 \enspace . 
\end{gather*}
Proceed to $\tilde p:=\prod_{\bar\jmath} r_{\bar\jmath}\cdot
\sum_{\bar\imath} \frac{q_{\bar\imath}}{r_{\bar\imath}}\cdot 
 Y^{\bar\imath}$: This polynomial in $E[X_1,\ldots,X_n,Y_1,\ldots,Y_m]$
is non-zero (e.g. on $x_1,\ldots,x_n$) and vanishes on 
$x_1,\ldots,x_n,y_1,\ldots,y_m\in\calX\cup\calY$.

Conversely let $\calX\cup\calY$ be algebraically dependent over $E$.
Then it holds $p(x_1,\ldots,x_n,y_1,\ldots,y_m)=0$
for some $n,m\in\IN$, $x_1,\ldots,x_n\in X$, $y_1,\ldots,y_m\in Y$,
and non-zero $p\in E[X_1,\ldots,X_n,Y_1,\ldots,Y_m]$.
A fortiori, $q:=p(x_1,\ldots,x_n,\cdots)\in E(X)[Y_1,\ldots,Y_m]$
satisfies $q(y_1,\ldots,y_m)=0$. To conclude algebraic independence
of $y_1,\ldots,y_m$ over $E(X)$, it remains to show $q\not=0$.
$0\not=p\in E[X_1,\ldots,X_n,Y_1,\ldots,Y_m]$ implies that
there exist $z_1,\ldots,z_m\in E$ such that
$0\not=p(X_1,\ldots,X_n,z_1,\ldots,z_m)=r(X_1,\ldots,X_n)\in E[X_1,\ldots,X_n]$.
Then $q(z_1,\ldots,z_m)=r(x_1,\ldots,x_n)\not=0$ holds
because $x_1,\ldots,x_n\in\calX$ are algebraically independent by hypothesis.
\item[c)]
Take some hypothetical non-trivial factorization
$p=q_1\cdot q_2$ in $E[X_1,\ldots,X_n,Y_1,\ldots,Y_m]$.
A fortiori, $p(\vec x,\vec Y)=q_1(\vec x,\vec Y)\cdot q_2(\vec x,\vec Y)$
constitutes a factorization in $E(\vec x)[\vec Y]$;
a non-trivial one: because if for instance 
$q_1(\vec x,\vec Y)$ were the constant polynomial,
say $q_1(\vec x,\vec Y)=c\in E$,
then $q_1(\vec X,\vec y)-c\not=0$ for some $y_1,\ldots,y_m\in E$
(since $q_1$ is by presumption a non-trivial factor of $p$)
constitutes a non-zero polynomial in $E[\vec X]$
vanishing on $x_1,\ldots,x_n$: contradicting that the latter
are algebraically independent over $E$.

Conversely suppose
$p(\vec x,\vec Y)=
q_1(\vec x,\vec Y)\cdot q_2(\vec x,\vec Y)$
in $E(\vec x)[\vec Y]$ and consider the polynomial
$r:=p-q_1\cdot q_2\in E[\vec X,\vec Y]$.
Although vanishing on $(\vec x,\vec Y)$,
it cannot be identically zero because that would
mean a non-trivial factorization of irreducible $p$.
On the other hand $r(\vec X,y_1,\ldots,y_m)\not=0$
for some $y_1,\ldots,y_m\in E$ would constitute a
non-zero polynomial in $E[\vec X]$ vanishing
on $x_1,\ldots,x_n$: contradicting that the latter
are algebraically independent over $E$.
\item[d)]
Since $(\vec x,\vec y)$ are algebraically independent over $E$,
$p(\vec x,\vec y,Z)$ is irreducible in $E(\vec x,\vec y)[Z]$ by c).
Since $(\vec y,z)$ are algebraically dependent over $E$,
$q(\vec y,z)=0$ for some non-zero $q\in E[\vec Y,Z]$;
w.l.o.g., $q$ is irreducible: and so is $q(\vec y,Z)$
in $E(\vec x,\vec y)[Z]$, again by c).
Each $p(\vec x,\vec y,Z)$ and $q(\vec y,Z)$ vanishes on $z$,
hence they share a common factor $r\in E(\vec x,\vec y)[Z]$;
but both being irreducible requires that they all coincide.
\qed\end{enumerate}
\end{proof}
\begin{proposition}
For any fixed $\vec z\in\IR^*$, 
$\IT$ is \BCSS decidable relative to
$\II_{\vec z}$; which is in turn decidable
relative to $\II:=\II_{()}$.
In formula: ~$\IT\BssReduceq\II_{\vec z}\BssReduceq\II$.
\end{proposition}
\begin{proof}
Suppose we are given oracle access to $\II$.
Since $\vec z$ is fixed, a \BCSS machine may store as constants
a transcendence basis $\vec y$ for $\IQ(\vec z)$ over $\IQ$.
Given $\vec x\in\IR^*$, it can then decide membership to
$\II_{\vec z}$ by querying ``$(\vec y,\vec x)\in\II$?'':
Since $\vec y$ is algebraically independent over $\IQ$
by construction, $(\vec y,\vec x)$ is iff $\vec x$ is
over $\IQ(\vec y)$ (Lemma~\ref{l:Indep}b)
or, equivalently, over $\IQ(\vec z)$.

Conversely given $x$, query membership to 
$\II_{\vec z}\supseteq\IT$ 
and accept if the answer is positive.
Otherwise $(\vec y,x)$ is algebraically dependent over $\IQ$,
hence there exists for some non-zero 
polynomial $p\in\IZ[\vec Y,X]$ irreducible
over $\IQ[\vec Y,X]$
and vanishing on $(\vec y,x)$.
Moreover such $p$ can be sought for (and hence found):
By the Gau\ss{} Lemma \mycite{Theorem~IV.\textsection 2.3}{Lang},
$p\in\IZ[\vec Y,X]$
is irreducible over $\IQ[\vec Y,X]$
iff it is irreducible over $\IZ[\vec Y,X]$;
and the latter property is decidable by
testing the finitely many candidate divisors
$q\in\IZ[\vec Y,X]$ of $\deg_{i}(q)\leq\deg_{i}(p)$
whose coefficients $q_i\in\IZ$ 
divide $p_i$ for all $i$.
Now once such $p=p(\vec Y,X)$ is found,
check whether it actually `depends' on 
(i.e. has in dense representation a nonzero 
coefficient to) some $Y_i$:
According to Lemma~\ref{l:Indep}d),
this is the case ~iff~ $x$ is
transcendental over $\IQ$.
\qed\end{proof}

\section{Real Incompressibility Method}
Discrete Kolmogorov Complexity Theory is a useful tool for establishing
(lower and average) bounds on running times of specific algorithms as well as generally
on the complexity of certain problems \mycite{Section~6}{Vitanyi}.
The same can be said about its \BCSS counterpart \mycite{Corollary~4}{Montana}.
For instance we conclude from Example~\ref{x:Transc}a) 
an entirely new proof of the following 

\begin{observation} \label{o:Pairing}
There exists no \BCSS-computable surjective 
(and in particular no fully real pairing) function 
$f:\IR\to\IR\times\IR$.
\end{observation}
\begin{proof}
Suppose that $f$ is computable by machine $\IM$
with constants $c_1,\ldots,c_J$.
Iteration yields a 
surjection $f^{(n)}:\IR\to\IR^n$ for any fixed $n$,
computable again by a machine with constants $c_1,\ldots,c_J$.
Take $n\in\IN$ and $\vec z\in\IR^n$ of Kolmogorov
Complexity much larger than $J$
according to Example~\ref{x:Transc}a).
By surjectivity, there exists $\zeta\in\IR$ with 
$f^{(n)}(\zeta)=\vec z$. Thus, $\vec z$ can be output
by storing the single 
constant $\zeta$ and invoking the machine evaluating $f^{(n)}$:
contradicting $\KOL(\zeta)\approx J\ll\KOL(\vec z)$.
\qed\end{proof}
%

\section{Miscellaneous}
This section handles off few further, related topics from classical 
computability theory \mycite{Section~5.6}{Moret}
(see also \cite{Moss})
in the context of real number computation:
\person{Rad\'{o}}'s \textsf{Busy Beaver} function,
\textsf{Quines}, and \person{Kleene}'s \textsf{Recursion}
and \textsf{Fixed Point Theorems}.
\subsection{Busy Beaver}
Classically, the busy beaver function $\Sigma(n)$ amounts to
the length of a longest string $\bar x\in\{0,1\}^*$ output by a terminating,
input-free Turing machine $M$ of $\length(\langle M\rangle)\leq n$.
It is well-known, as is the Kolmogorov complexity function, incomputable,
approximable, and equivalent to the Halting problem.

Now every Turing machine $M$ can be simulated by a \BCSS machine
$\IM$ of $\size(\langle\IM\rangle)=1$ independent of $\length(\langle M\rangle)$;
hence it does \emph{not} make sense to ask the following
\begin{question}[unreasonable]
What is the maximum size of a string $\vec x\in\IR^*$ output by a
terminating, input-free \BCSS machine $\IM$ of $\size(\langle\IM\rangle)\leq n$ ?
\end{question}
The answer is, of course: infinite.

In view of Theorem~\ref{t:Transc}, one might be tempted to instead consider
\begin{question}
What is the maximum transcendence degree of a string $\vec x\in\IR^*$
output by a 
terminating, input-free \BCSS machine $\IM$ of $\size(\langle\IM\rangle)\leq n$ ?
\end{question}
However, again, this question is easy to answer (namely ``$n$'')
and to compute.
\subsection{Quines, Fixed-point and Recursion Theorems}
A quine is a program $p$ which (upon empty input) outputs itself (e.g. its own source code)
and terminates. More generally, one may demand that $p$ performs
some prescribed computable operation on its input $x$ and on 
its own encoding which, however, is \emph{not} passed as input.
Solutions to both problems are well-known to exist in the
discrete realm and amount to Kleene's first and second
Recursion Theorem, respectively. Closely related is his
Fixed Point Theorem, asserting that every recursive total
function on G\"{o}del indices has a (semantic) fixed point.

All of them immediately carry over to the
real setting: Since a \BCSS machine $\IM$ accesses its constants
by reference, it suffices to consider only $\IM$'s finite control
$\delta$ --- to which the discrete theorems apply.
Alternatively, their classical proofs based on SMN
and UTM properties translate literally to the
real setting (recall Section~\ref{s:BSS}a).

\begin{observation}
Fix a universal \BCSS machine $\IU$.
\begin{enumerate}
\item[a)]
To any \BCSS machine $\IM$ (with constants $c_1,\ldots,c_J$), 
there exists another one $\IM'$ (again with constants $c_1,\ldots,c_J$)
such that $\IM'$ on $\vec x$ behaves like $\IM$ on $(\langle\IM\rangle,\vec x)$.
\item[b)]
To every total \BCSS-computable function $f:\IR^*\to\IR^*$,
there exists some $\vec x\in\IR^*$ such that
\begin{equation} \label{e:Fixedpoint}
\forall \vec y\in\IR^*: \quad 
\IU\big(\vec x,\vec y\big)\;=\;\IU\big(f(\vec x),\vec y\big) \enspace .
\end{equation}
Moreover, if $f$ is realized by $\IM$,
the mapping $\langle\IM\rangle\to\vec x$ is \BCSS-computable.
\end{enumerate}
\end{observation}
The equality in (\ref{e:Fixedpoint}) is meant in the extended sense that
either side is undefined iff the other is.

\section{Conclusion}
The present work has extended the work \cite{Montana}
and its real variant of Kolmogorov complexity theory.
Some important properties have turned out to carry over,
however with considerably different proofs.
Specifically, `most' real vectors have complexity equal to their
length; and the complexity of a given string
can be computationally approximated from above
but not determined exactly.
However opposed to the classical discrete case,
real Kolmogorov Complexity is not reducible
\emph{from} the real Halting problem $\IH$.

We close with some open
\begin{question}
\begin{enumerate}
\item[a)]
  Does Proposition~\ref{p:Approx} extend to $\KOLO_{\vec z}$? \\
  Does Theorem~\ref{t:Incomplete} extend to 
  $\ISO_{\vec z}:=\{(\vec x,k):\KOLO_{\vec z}(\vec x)\leq k\}\subseteq\IR^*\times\IN$ ?
\item[b)]
  Theorem~\ref{t:Uncomputable} is only concerned with \BCSS G\"{o}delizations
  induced by machines of the form $\IU_{\vec z}$. Does it extend to all 
  universal machines $\IU$?
\item[c)]
  How about the complex case, i.e. w.r.t. \BSS-machines 
  over $\IC$ permitted tests only for equality?
\end{enumerate}
\end{question}

\paragraph{Acknowledgments:}
The first author is grateful to his colleagues 
\person{Dennis Amelunxen}, 
\person{Peter Scheiblechner}, and \person{Thorsten Wedhorn}
for discussions about algebraic varieties
and the rationality questions arisen in Section~\ref{s:Segre}.
Moreover, \person{Peter Scheiblechner} has been a great help
in finding a proof for Lemma~\ref{l:Indep}c).
Finally we owe to \person{Klaus Meer} for pointing us to the
seminal work of \person{Monta\~{n}a} and \person{Pardo}
who first introduced real Kolmogorov Complexity.

\begin{appendix}
\section{Facsimile of \cite{Segre}}
\includegraphics[width=\textwidth]{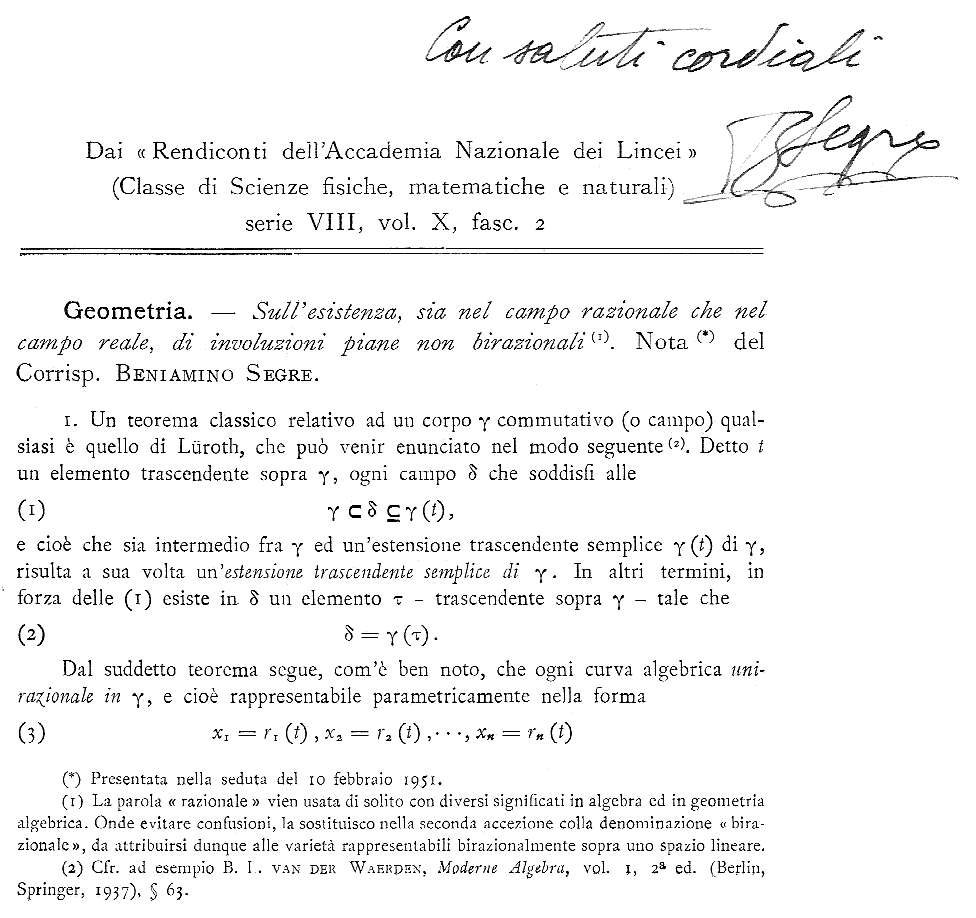}

\noindent
\includegraphics[width=\textwidth]{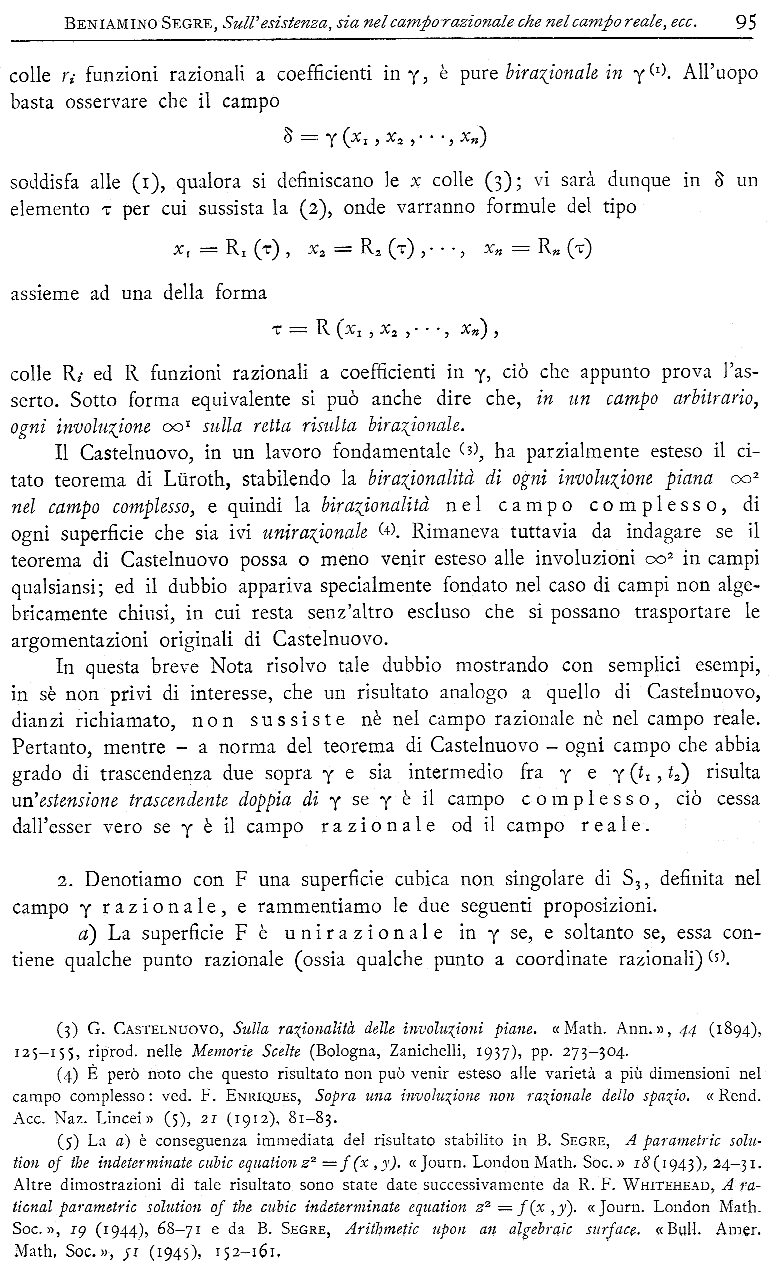}

\noindent
\pagestyle{empty}
\includegraphics[width=\textwidth]{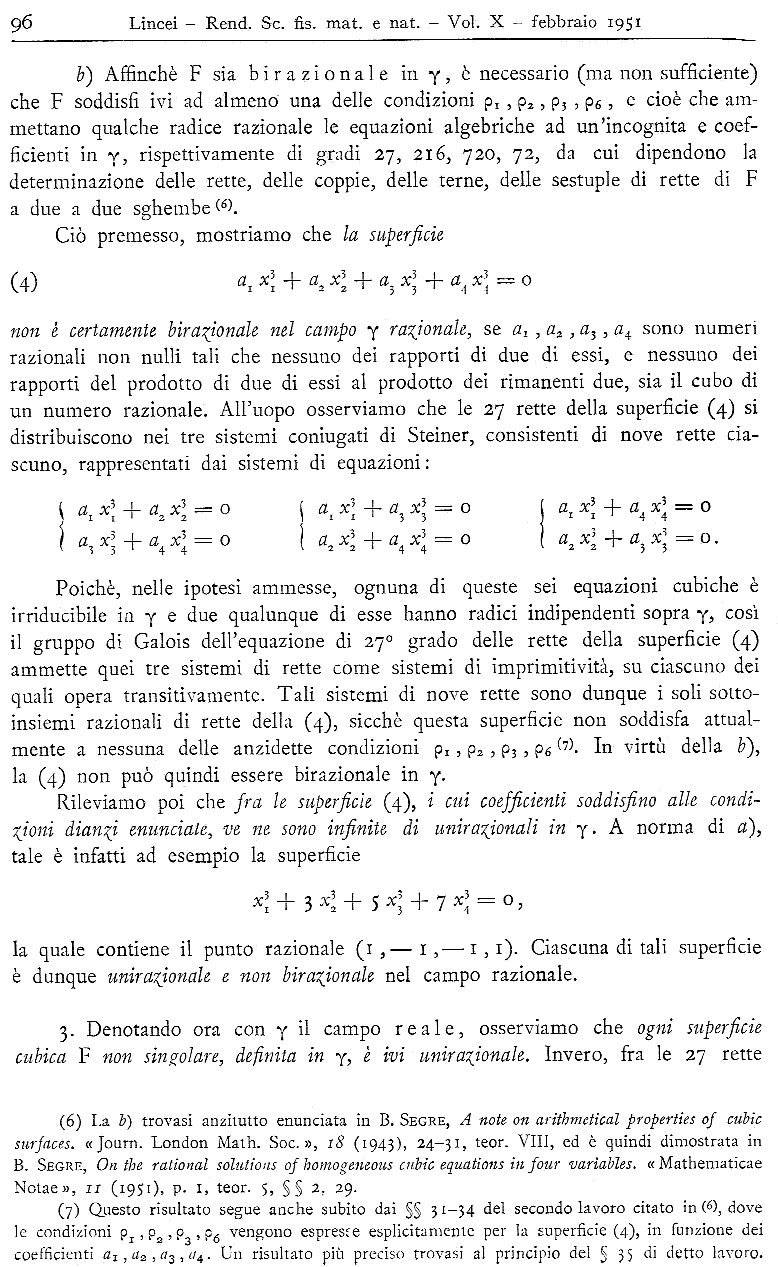}

\noindent
\includegraphics[width=\textwidth]{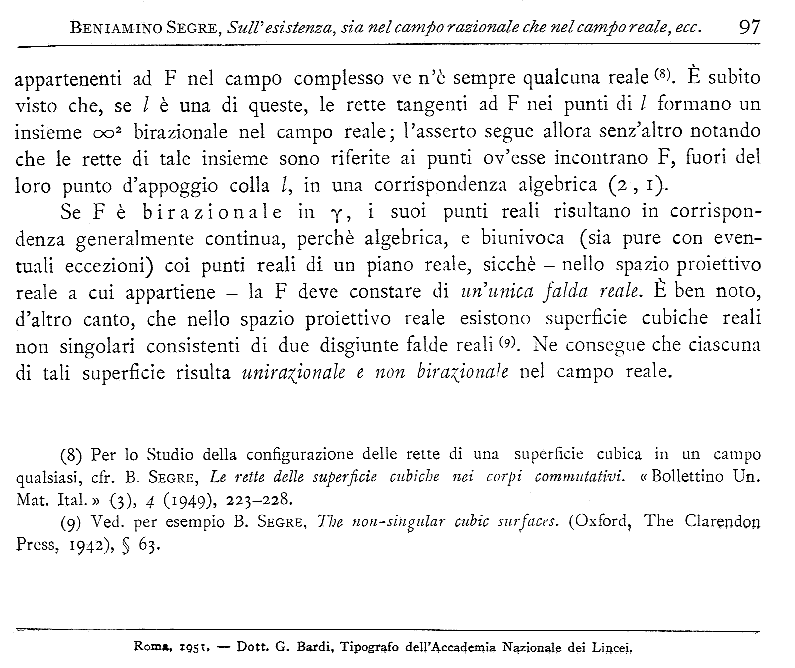}
\end{appendix}
\pagestyle{empty}
\end{document}